\documentclass[prd,aps,floats,floatfix,eqsecnum,nofootinbib]{revtex4}
\usepackage{graphics}
\usepackage{graphicx}
\usepackage{rotate}
\usepackage{color}
\usepackage{rotating}
\usepackage{psfrag}
\usepackage{amsmath,amsthm}
\usepackage{epsfig}
\usepackage{slashed}
\usepackage{sidecap}

\begin{document}
\newcommand{\D}{\displaystyle} 
\newcommand{\T}{\textstyle} 
\newcommand{\SC}{\scriptstyle} 
\newcommand{\SSC}{\scriptscriptstyle} 
\newcommand{\be}{\begin{equation}}
\newcommand{\ee}{\end{equation}}
\newcommand{\avg}[1]{\langle #1 \rangle}
\newcommand{\vx}{{\boldsymbol{x}}}
\newcommand{\rv}{{\boldsymbol{r}}}
\newcommand{\vq}{\ensuremath{\vec{q}}}
\newcommand{\pv}{\ensuremath{\vec{p}}}
\def\AJ{{\it Astron. J.} }
\def\ARAA{{\it Annual Rev. of Astron. \& Astrophys.} }
\def\ApJ{{\it Astrophys. J.} }
\def\ApJL{{\it Astrophys. J. Letters} }
\def\ApJS{{\it Astrophys. J. Suppl.} }
\def\ApP{{\it Astropart. Phys.} }
\def\AA{{\it Astron. \& Astroph.} }
\def\AAR{{\it Astron. \& Astroph. Rev.} }
\def\AAL{{\it Astron. \& Astroph. Letters} }
\def\AASu{{\it Astron. \& Astroph. Suppl.} }
\def\AN{{\it Astron. Nachr.} }
\def\IJMP{{\it Int. J. of Mod. Phys.} }
\def\JGR{{\it Journ. of Geophys. Res.}}
\def\JHEP{{\it Journ. of High En. Phys.} }
\def\JPhG{{\it Journ. of Physics} {\bf G} }
\def\MNRAS{{\it Month. Not. Roy. Astr. Soc.} }
\def\Nature{{\it Nature} }
\def\NewAR{{\it New Astron. Rev.} }
\def\NJPh{{\it New Journ. of Phys.} }
\def\PASP{{\it Publ. Astron. Soc. Pac.} }
\def\PhFl{{\it Phys. of Fluids} }
\def\PLB{{\it Phys. Lett.}{\bf B} }
\def\PhysRep{{\it Phys. Rep.} }
\def\PR{{\it Phys. Rev.} }
\def\PRD{{\it Phys. Rev.} {\bf D} }
\def\PRL{{\it Phys. Rev. Letters} }
\def\RMP{{\it Rev. Mod. Phys.} }
\def\Science{{\it Science} }
\def\ZfA{{\it Zeitschr. f{\"u}r Astrophys.} }
\def\ZfN{{\it Zeitschr. f{\"u}r Naturforsch.} }
\def\etal{{\it et al.}}
\hyphenation{mono-chro-matic sour-ces Wein-berg
chang-es Strah-lung dis-tri-bu-tion com-po-si-tion elec-tro-mag-ne-tic
ex-tra-galactic ap-prox-i-ma-tion nu-cle-o-syn-the-sis re-spec-tive-ly
su-per-nova su-per-novae su-per-nova-shocks con-vec-tive down-wards
es-ti-ma-ted frag-ments grav-i-ta-tion-al-ly el-e-ments me-di-um
ob-ser-va-tions tur-bul-ence sec-ond-ary in-ter-action
in-ter-stellar spall-ation ar-gu-ment de-pen-dence sig-nif-i-cant-ly
in-flu-enc-ed par-ti-cle sim-plic-i-ty nu-cle-ar smash-es iso-topes
in-ject-ed in-di-vid-u-al nor-mal-iza-tion lon-ger con-stant
sta-tion-ary sta-tion-ar-i-ty spec-trum pro-por-tion-al cos-mic
re-turn ob-ser-va-tion-al es-ti-mate switch-over grav-i-ta-tion-al
super-galactic com-po-nent com-po-nents prob-a-bly cos-mo-log-ical-ly
Kron-berg Berk-huij-sen}

\title{\Large HIGHLIGHTS and CONCLUSIONS 

\medskip

of the Chalonge 13th Paris Cosmology Colloquium:

\medskip

'The Standard Model of the Universe: From Inflation to Today Dark Energy',    

\medskip

Ecole Internationale d'Astrophysique Daniel Chalonge,

Observatoire de Paris

in the historic Perrault building, July 2009.}

\author{\Large \bf   H. J. de Vega$^{(a,b)}$,  
M. C. Falvella$^{(c)  }$, N. G. Sanchez$^{(b)}$}

\date{\today}

\affiliation{$^{(a)}$ LPTHE, Universit\'e
Pierre et Marie Curie (Paris VI) et Denis Diderot (Paris VII),
Laboratoire Associ\'e au CNRS UMR 7589, Tour 24, 5\`eme. \'etage, 
Boite 126, 4, Place Jussieu, 75252 Paris, Cedex 05, France. \\
$^{(b)}$ Observatoire de Paris,
LERMA. Laboratoire Associ\'e au CNRS UMR 8112.
 \\61, Avenue de l'Observatoire, 75014 Paris, France.\\
$^{(c)}$ Italian Space Agency and MIUR, Viale Liegi n.26,
00198 Rome, Italy.}

\maketitle

\tableofcontents

\begin{figure}[htbp]
\epsfig{file=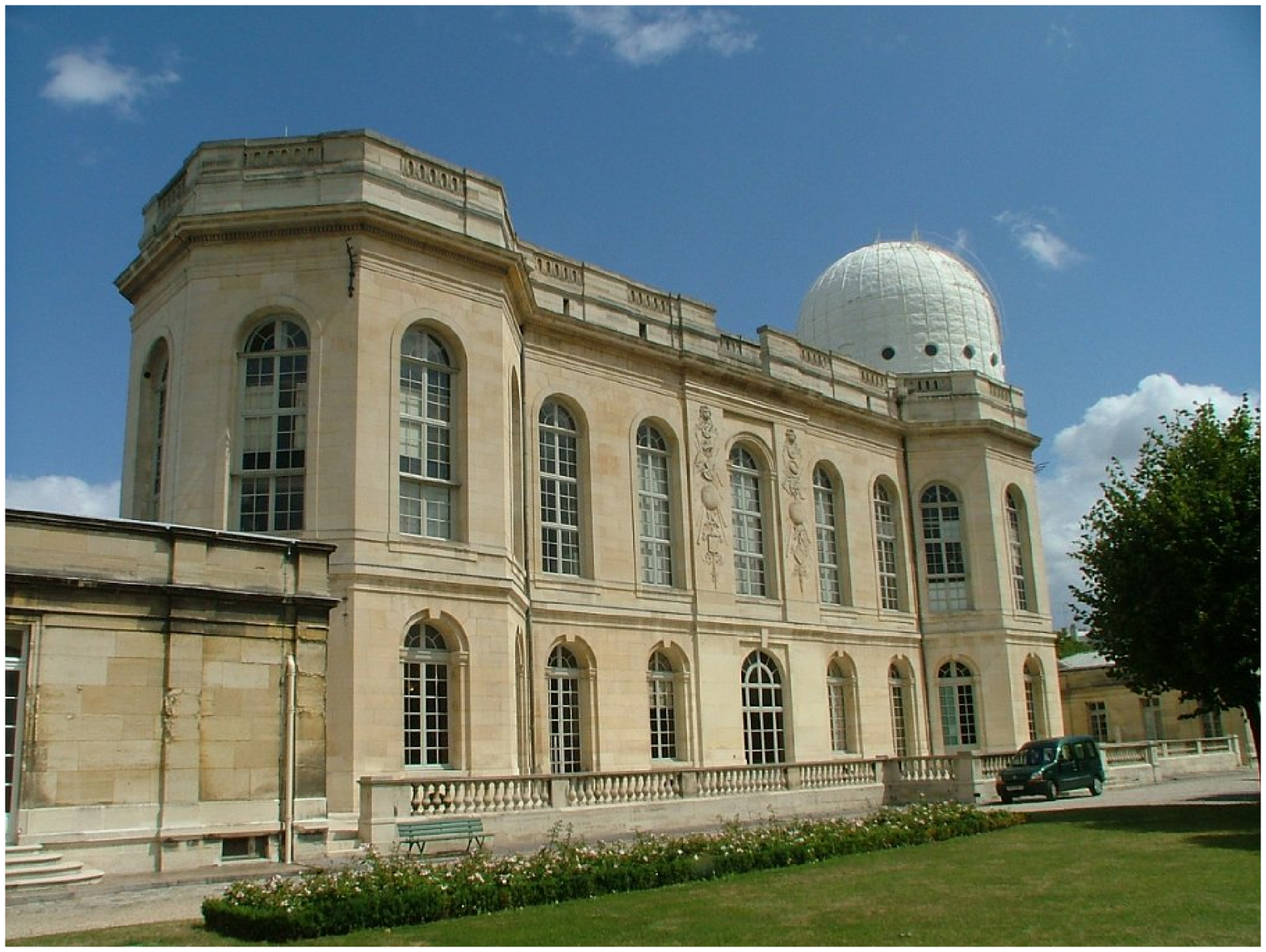,width=16cm,height=13cm}
\end{figure}

\begin{figure}[htbp]
\epsfig{file=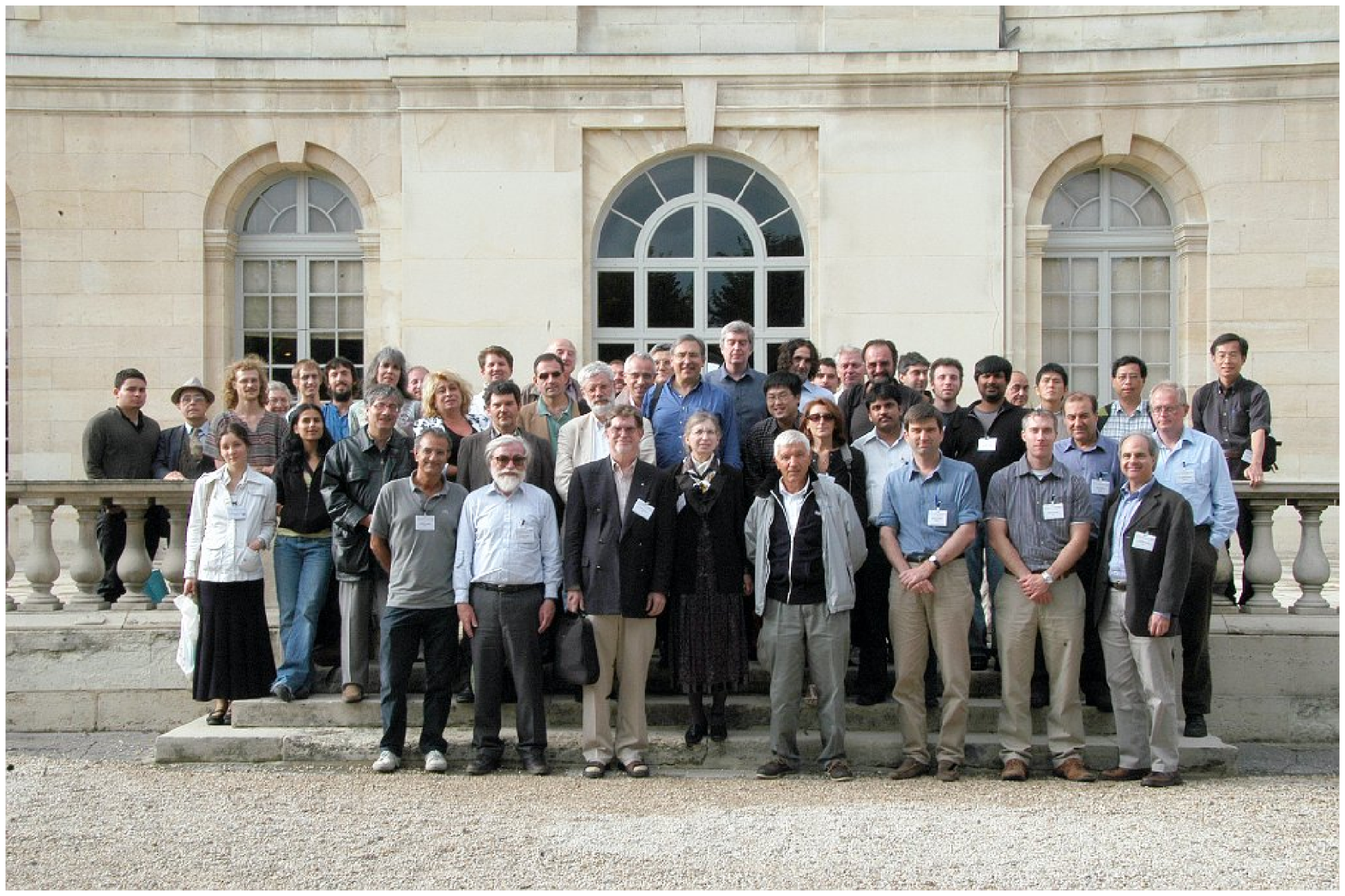,width=16cm,height=12cm}
\end{figure}

\newpage

\section{Purpose of the Colloquium and Introduction}

The main aim of the series "Paris Cosmology Colloquia", in the framework of the International School of Astrophysics  {\bf "Daniel Chalonge"},  is to put together real cosmological and astrophysical  data and hard theory approach connected to them. The Paris Cosmology Colloquia bring together physicists, astrophysicists and astronomers from the world over. Each year these Colloquia are more attended and appreciated both by PhD students, post-docs and lecturers. The format of the Colloquia is intended to allow easy and fruitful mutual contacts and communication.

\bigskip

The subject of the 13th Paris Cosmology Colloquium 2009  was "THE STANDARD MODEL OF THE UNIVERSE: FROM INFLATION TO TODAY DARK ENERGY", George Smoot, Nobel Prize of Physics 2006 and Daniel Chalonge Medal . 

\bigskip

The  Colloquium took  place during full three days  (Thursday July 23, Friday 24 and Saturday July 25) at the parisian campus of  Paris Observatory (HQ), in the historic Perrault building.

\bigskip

The {\bf 13th Paris Cosmology Colloquium 2009} was within the astrofundamental physics spirit of the Chalonge School, focalized on recent observational and theoretical progress on the CMB and inflation with predictive power, dark matter, dark energy, dark ages and LSS in the context of the Standard Model of the Universe. Never as in this period, the Golden Age of Cosmology, the major subjects of the Daniel Chalonge School were so timely and in full development: the WMAP mission released in April 2008 the new survey (5 years of observations) and the PLANCK mission has been launched (May 2009) and is performing its First Survey.

\bigskip

The {\bf main topics} were: Observational and theoretical progress in deciphering the nature of dark matter, dark energy, dark ages and the 21 cm line. Large and small scale structure formation. Inflation after WMAP (in connection with the CMB and LSS data), slow roll and 
fast roll inflation, quadrupole suppression and initial conditions; quantum effects. CMB polarization, primordial magnetic fields effects. Neutrinos in cosmology. Measurements of the CMB by the Planck mission and its science perspectives. 

\bigskip

All Lectures  are  plenary and followed by a discussion. 
Enough time is provided to the discussions.  

\begin{center}

Informations of the Colloquium are available on

\medskip

 {\bf http://chalonge.obspm.fr/colloque2009.html}

\end {center}

\bigskip

Informations on the previous Paris Cosmology Colloquia and  on the Chalonge school 
events are available at  

\begin{center}

 {\bf http://chalonge.obspm.fr}

\medskip

(lecturers, lists of participants, lecture files and photos during the Colloquia).

\end {center}

\bigskip

This Paris Colloquia series started in 1994 at the Observatoire de Paris. The series cover selected topics of high current interest in the interplay between cosmology and fundamental physics. The PARIS COSMOLOGY COLLOQUIA are informal meetings. Their purpose is an updated understanding, from a fundamental point of view, of the progress and current problems in the early universe, cosmic microwave background radiation, large scale structure and neutrinos in astrophysics and the interplay between them. Emphasis is given to the mutual impact of fundamental physics and cosmology, both at theoretical and experimental -or observational- levels. 

\bigskip

Deep understanding, clarification, synthesis, a careful interdisciplinarity within a fundamental physics approach, are goals of this series of Colloquia.

\bigskip

Sessions last for three full days and leave enough time for private discussions and to enjoy the beautiful parisian campus of Observatoire de Paris (built on orders from Colbert and to plans by Claude Perrault from 1667 to 1672).

\bigskip

Sessions take  place in the Cassini Hall, on the meridean of Paris, in "Salle du Conseil" (Council Room) in the historic Perrault building ("Bâtiment Perrault") of Observatoire de Paris HQ, under the portraits of Laplace, Le Verrier, Lalande, Arago, Delambre and Louis XIV and in the "Grande Galerie" (the Great Gallery).

\bigskip

An {\bf Exhibition} retraced the 18 years of activity of the Chalonge School and of George Smoot participation to the School along these 18 years.
The books and proceedings of the School since its creation, as well as historic Daniel Chalonge material, and Chalonge instruments were on exhibition at the Great Gallery.

\bigskip

During the Colloquium, the International School of Astrophysics "Daniel Chalonge" has awarded the  {\bf Daniel Chalonge Medal 2009} to {\bf Prof. Peter Biermann} from MPI  Institut of Radioastronomie of Bonn (D) and University of Alabama Tuscaloosa (USA) and for his outstanding support and contributions to the Chalonge School.

\bigskip

After the Colloquium, a tour of the Perrault building took place guided by Professor Suzanne Debarbat around the subject  "From Hipparque to the Hipparcos satellite".

\bigskip

More information on the Colloquia of this series can be found in the Proceedings of the Colloquia from 1994 (H.J. de Vega and N. Sánchez, Editors) published by World Scientific Co. and by Observatoire de Paris.

\bigskip

We want to express our grateful thanks to all the sponsors of the Colloquium,  to all the lecturers for their excellent and polished presentations, to all the lecturers and participants for their active participation and their contribution to the outstanding discussions and lively atmosphere, to the assistants, secretaries and all collaborators of the Chalonge School,  who made this event so harmonious, wonderful and successful . 

\bigskip

\begin{center}
                                   
With Compliments and kind regards,

\bigskip  
               
\bigskip  
                             
{\bf Hector J de Vega, Maria Cristina Falvella,  Norma G Sanchez}

\end{center}

\begin{figure}[htbp]
\epsfig{file=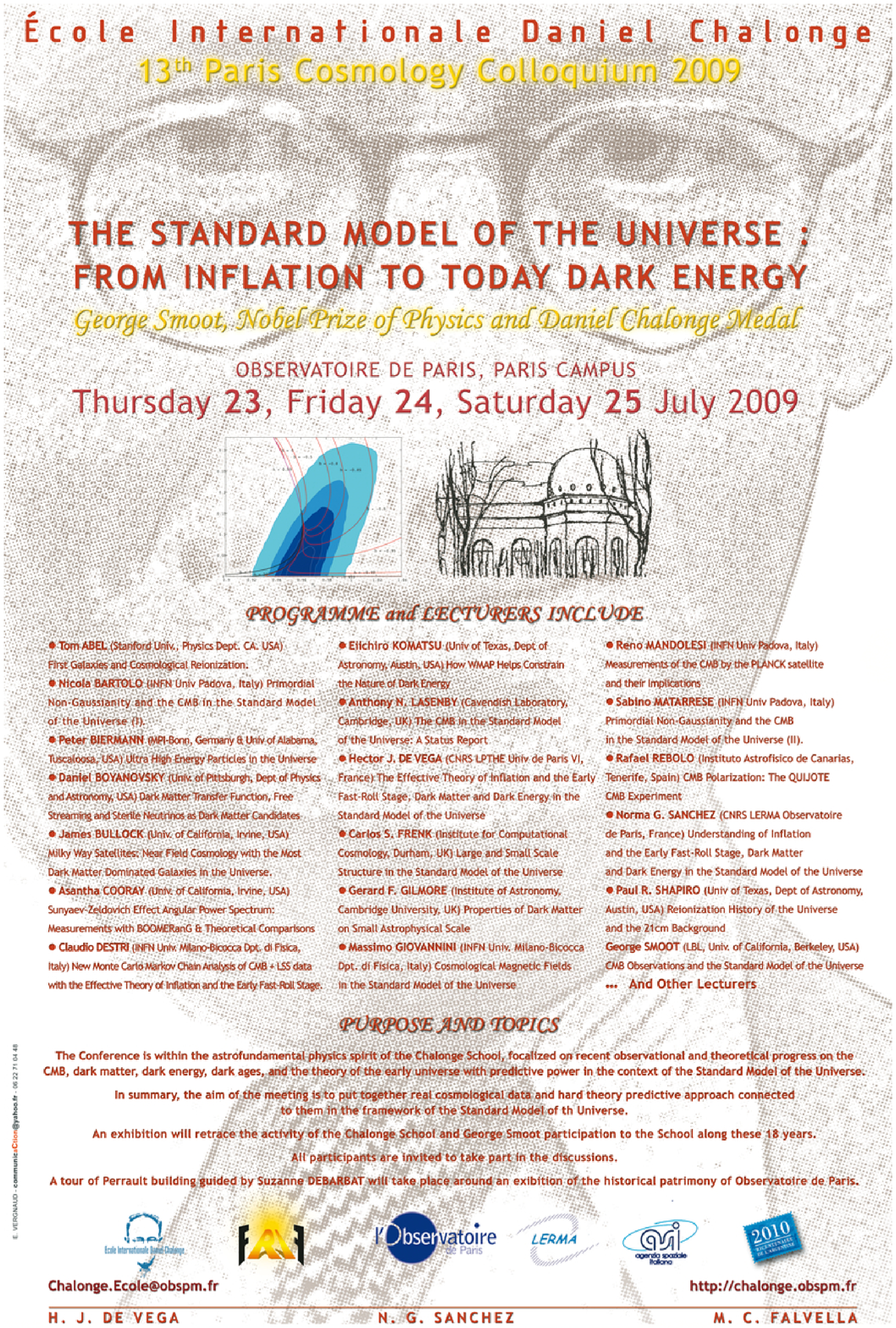,width=16cm,height=22cm}
\end{figure}

\newpage

\section{Programme and Lecturers}

\begin{itemize}

\item{{\bf Tom ABEL} (Stanford Univ.,  Physics Dept. CA, USA)
First Galaxies and Cosmological Reionization.}

\item{{\bf Nicola BARTOLO} (INFN Univ Padova, Italy)
Primordial Non-Gaussianity and the CMB in the Standard 
Model of the Universe (I.)}

\item{{\bf Peter BIERMANN}  (MPI-Bonn, Germany and Univ of Alabama, Tuscaloosa, USA)
Ultra High Energy Particles in the Universe}

\item{{\bf Daniel BOYANOVSKY} (Univ. of  Pittsburgh, Dept of Physics and Astronomy, USA)
Dark Matter Transfer Function, Free Streaming and Sterile Neutrinos as Dark Matter Candidates}

\item{{\bf James BULLOCK} (University of California, Irvine, USA)
Milky Way Satellites: Near Field Cosmology with the Most Dark Matter Dominated Galaxies in the Universe.}  

\item{{\bf Asantha COORAY} (University of California, Irvine, USA)
Cosmology with the  21 cm Background}

\item{{\bf Claudio DESTRI} (INFN Univ. Milano-Bicocca Dpt. di Fisica, Italy) 
New Monte Carlo Markov Chain Analysis of CMB +LSS  data 
with the Effective Theory of Inflation and the Early Fast-Roll Stage.}

\item{{\bf Hector J. DE VEGA} (CNRS LPTHE Univ de Paris VI, France)
The Effective Theory of Inflation and the Early Fast-Roll Stage, Dark Matter and Dark Energy in the Standard Model of the Universe}

\item{{\bf Carlos S. FRENK} (Institute for Computational Cosmology, Durham, UK)
The Small-Scale Structure of the Universe} 

\item{{\bf Gerard F. GILMORE}, (Institute of Astronomy, Cambridge University, UK)
Properties of Dark Matter on Small Astrophysical Scales}

\item{{\bf Massimo GIOVANNINI} (INFN Univ. Milano-Bicocca Dpt. di Fisica, Italy)
Cosmological Magnetic Fields in the Standard Model of the Universe}

\item{{\bf Alexander KASHLINSKY} (NASA Goddard Space Flight Center, Greenbelt, MD, USA) 
Probing large-scale peculiar flows of clusters of galaxies.}

\item{{\bf Eiichiro  KOMATSU} (Univ of Texas, Dept of Astronomy, Austin, USA) 
How WMAP Helps Constrain the Nature of Dark Energy}

\item{{\bf Anthony N. LASENBY} (Cavendish Laboratory, Cambridge, UK)
 The CMB in the Standard Model of the Universe: A Status Report}

\item{{\bf Reno MANDOLESI} (INAF-IASF Bologna, Italy): 
Measurements of the CMB by the PLANCK satellite and their Implications}

\item{{\bf Sabino MATARRESE} (INFN Univ Padova, Italy):       
Primordial Non-Gaussianity and the CMB in the Standard Model of the Universe (II).}

\item{{\bf Rafael REBOLO} (Instituto Astrofisico de Canarias, Tenerife, Spain)
CMB Polarization: The QUIJOTE CMB Experiment}

\item{{\bf Paolo   SALUCCI} (SISSA-Astrophysics, Trieste, Italy) 
 The Dark Matter Surface Density in Galaxies and Cored Density Profiles}

\item{{\bf Norma G. SANCHEZ} (CNRS LERMA Observatoire de Paris, France)
Understanding of Inflation and the Early Fast-Roll Stage, Dark Matter and Dark Energy in the Standard Model of the Universe}

\item{{\bf Paul R. SHAPIRO}, Univ of Texas, Dept of Astronomy, Austin, USA   
Reionization History of the Universe and the 21cm Background}

\item{{\bf George SMOOT} (LBL, Univ. of California, Berkeley, USA)
CMB Observations and the Standard Model of the Universe}

\end{itemize}

\newpage

\section{Highlights by the Lecturers}

\begin{center}
  
More  informations on the Colloquium Lectures  are at: 

{\bf http://www.chalonge.obspm.fr/colloque2009.html}

\end{center}

\subsection{Peter Biermann$^{1,2,3,4,5}$}

 {\bf with Julia Becker$^{6,7}$, Lauren\c{t}iu Caramete$^{1,8}$, Laszlo \'A. Gergely$^{9}$, Ioana C. Mari\c{s}$^{5}$,  Athina Meli$^{10}$, Eun-Suk Seo$^{11}$, Vitor de Souza$^{12}$, Todor Stanev$^{13}$, Oana Ta\c{s}c\u{a}u$^{14}$}\\

\begin{center}

$^1$ MPI for Radioastronomy, Bonn, Germany\\
$^2$ Dept. of Phys. \& Astron., Univ. of Bonn, Germany \\
$^3$ Dept. of Phys. \& Astr.,
Univ. of Alabama, Tuscaloosa, AL, USA\\
$^4$Dept. of Phys.,
Univ. of Alabama at Huntsville, AL, USA\\
$^{5}$ Inst. Nucl. Phys. FZ, Karlsruhe Inst. of Techn. (KIT), Germany\\
$^{6}$ Dept. of Phys., Univ. Bochum, Bochum, Germany\\
$^{7}$ Institution f{\"o}r Fysik, G{\"o}teborgs Univ., Sweden\\
$^{8}$ Institute for Space Studies, Bucharest, Romania \\
$^{9}$ Phys. Dept., Univ. of Szeged, Szeged, Hungary\\
$^{10}$ ECAP, Physik. Inst. Friedrich-Alexander
Univ. Erlangen-N{\"u}rnberg, Germany\\
$^{11}$ IPST and Dept. of Physics, Univ. of Maryland, College Park,
MD, USA\\
$^{12}$ Universidade de S$\tilde{a}$o Paulo, Instituto de F\'{\i}sica de
S$\tilde{a}$o Carlos, Brazil\\
$^{13}$ Bartol Research Inst., Univ. of Delaware, Newark, DE, USA \\
$^{14}$ Phys. Dept., Univ. Wuppertal, Germany\\

\bigskip

{\bf Ultra High Energy Particles and Cosmic Ray \\
Electrons/Positrons: from Massive Star Explosions}

\end{center}

\bigskip

The subtle properties of massive star are shown to be key to understand 
various new results in cosmic rays, both at low and at extremely high energies.

\medskip

The recent discovery of an excess cosmic ray electron and positron component is naturally explained in the context of recognizing, that stars with magnetic winds have a well-known topology in their wind: 

\medskip

This means (1993) that the cosmic rays resulting from the explosion of such stars have a small polar cap component with $E^{-2}$, and for most of their surface give a spectrum of $E^{-7/3}$.  This explains readily (2009) the new results from Pamela, ATIC, Fermi and H.E.S.S., using the model predicted in 1993. 

\medskip

Ultra high energy cosmic rays may be heavy nuclei, and then the question is from what source: The radio galaxy Cen A is nearby, and has a decaying star-burst, when very many massive stars were formed, which have recently exploded.  The relativistic shock in the jet in Cen A can only contain heavy nuclei at very high energies.  \\
In such a case, the spectrum of the various heavy elements from the polar cap component can be boosted up in energy by $\Gamma_{sh}^{2} = 2500$, possibly explaining the spectrum of these high energy particles and their chemical composition.  Many more predictions follow from such a scheme, which therefore can easily be disproven, or supported.

\medskip

PARTICLES AS TRACERS FOR THE MOST MASSIVE EXPLOSIONS IN THE MILKY WAY

\medskip

At their deaths, the most massive stars in the Milky Way seem to leave
behind unambiguous signatures of particles. Recently, a 
population of electrons and positrons was observed by several
experiments. Attempts were made to attribute such a signal to the decay
of dark matter.  However, a natural explanation is in the explosions of
giant stars that are more than 15 times heavier than our sun. 

\medskip

A dying star of very high mass ejects most of its matter in a final explosion,
which then ploughs its way through a massive stellar wind. During this
process, electrons and positrons escape from two different regions: the
lower energy signal comes from the entire surface of the exploding star,
while at higher energies, the regions around the poles of the rotating
star start to dominate. This naturally explains the observed energy
behavior of electrons and positrons.

\medskip

{\bf References:}

"`Cosmic ray electrons and positrons from supernova explosions of massive
stars"', 
P.L. Biermann, J.K. Becker, A. Meli, W. Rhode, E.-S. Seo and T. Stanev

Phys.Rev.Lett.103:061101,2009,      
arXiv:0903.4048 Categories: astro-ph.HE astro-ph.GA

\newpage

\subsection{James Bullock}

\vskip -0.3cm

\begin{center}

Center for Cosmology, University of California, Irvine, USA.

\bigskip

{\bf Dark Matter and Dwarf Galaxies: Evidence for a  threshold mass in galaxy formation?} 

\end{center}

\medskip

\begin{itemize}

\item{ We have derived a new and accurate mass-estimator for dispersion supported galaxies that is correct for general assumptions about stellar velocity anisotropy and dark matter vs. stellar content.  Specifically the mass within the $3-d $ half-light radius $r_{1/2}$ of a stellar system is given by the following simple, yet accurate formula:  $M(r_1/2) = 3 G^{-1} r_{1/2} \sigma_{los}^2$, where $\sigma_{los}$ is the luminosity-weighted line-of-sight velocity dispersion.\\  
Reference: Wolf et al. (2009, to be submitted)}

\bigskip

\item{ All of the dwarf satellite galaxies of the Milky Way are consistent with inhabiting a halo of a common mass, $M_{vir} ~ 10^9 Msun$.  Remarkably, the least luminous dwarfs, with luminosities as low as $300 L_{sun}$ seem to inhabit dark matter halos that are just as massive as those of their more luminous counterparts, which are $~ 10,000$ times brighter.  The lack of observed trend between central dark matter density and luminosity is difficult to explain with current models and may be indicative of a low-mass threshold in galaxy formation. \\ 
{\bf Reference}: Strigari et al. (2008, Nature)}

\bigskip

\item{ We have used completeness limits from the SDSS to argue that there is likely a very large population of undiscovered, low-luminosity dwarf galaxies orbiting within the halo of the Milky Way.  Straightforward corrections indicate that there are approximately $~500$ galaxies with luminosities greater than $1000 L_{sun}$ within $400 kpc$ of the Sun.  Future surveys like LSST can detect these galaxies.  Whether they are discovered or not, these searches will provide important constraints on the nature of dark matter and on models of galaxy formation.\\
{\bf Reference}: Tollerud et al. (2008, ApJ)}

\bigskip

\item{ Dwarf satellite galaxies provide ideal astrophysical sources for dark matter indirect detection experiments because they have high dark matter densities, neglibigle astrophysical backgrounds, and are fairly nearby.  Sculptor and Segue 1 are the most promising candidates for Fermi and ACTs.\\
{\bf Reference}: Strigari et al. (2008, ApJ) + Martinez et al. (2009, JCAP)}

\end{itemize}

\newpage

\subsection{C. Destri, Hector J. de Vega and N.G. Sanchez}

\vskip -0.3cm

\begin{center}

C.D:  INFN/Univ Milano-Bicocca, Dipt di Fisica G. Occhialini, Milano, Italia\\
H.J.dV: LPTHE, CNRS/Universit\'e Paris VI-P. \& M. Curie \& Observatoire de Paris, Paris, France\\
N.G.S: LERMA, CNRS/Observatoire de Paris, Paris, France

\bigskip

{\bf The Effective Theory of Inflation in the Standard Model of the Universe and the CMB+LSS data analysis}

\end {center}

\bigskip

Inflation is today a part of the Standard Model of the Universe supported by
the cosmic microwave background (CMB) and large scale structure (LSS)
datasets.

\bigskip

Inflation solves the horizon and flatness problems and
naturally generates  density fluctuations that seed LSS and CMB anisotropies,
and tensor perturbations (primordial gravitational waves). 

\bigskip

Inflation theory is
based on a scalar field  $ \varphi $ (the inflaton) whose potential
is fairly flat leading to a slow-roll evolution. 

\bigskip

This review focuses on the following new aspects of inflation. We present the
effective theory of inflation \`a la {\bf Ginsburg-Landau} in which
the inflaton potential is a polynomial in the field $ \varphi $ and has
the universal form $ V(\varphi) = N \; M^4 \;
w(\varphi/[\sqrt{N}\; M_{Pl}]) $, where $ w = {\cal O}(1) , \;
M \ll M_{Pl} $ is the scale of inflation and  $ N \sim 60 $ is the number
of efolds since the cosmologically relevant modes exit the horizon till
inflation ends.

\bigskip

The slow-roll expansion becomes a systematic $ 1/N $ expansion and
the inflaton couplings become {\bf naturally small} as powers of the ratio
$ (M / M_{Pl})^2 $. The spectral index and the ratio of tensor/scalar
fluctuations are $ n_s - 1 = {\cal O}(1/N), \; r = {\cal O}(1/N) $ while
the running index turns to be $ d n_s/d \ln k =  {\cal O}(1/N^2) $
and therefore can be neglected. The {\bf energy scale of inflation }$ M \sim 0.7
\times 10^{16}$ GeV is
completely determined by the amplitude of the scalar adiabatic
fluctuations. 

\bigskip

A complete analytic study plus the
Monte Carlo Markov Chains (MCMC) analysis of the available
CMB+LSS data (including WMAP5) with fourth degree trinomial potentials
showed: 

\begin {itemize}

\item {{\bf(a)} the {\bf spontaneous breaking} of the
$ \varphi \to - \varphi $ symmetry of the inflaton potential.} 

\medskip

\item {{\bf(b)} a {\bf lower bound} for $ r $ in new inflation:
$ r > 0.023 \; (95\% \; {\rm CL}) $ and $ r > 0.046 \;  (68\% \;
{\rm CL}) $. }

\medskip

\item {{\bf(c)} The preferred inflation potential is a {\bf double
well}, even function of the field with a moderate quartic coupling
yielding as most probable values: $ n_s \simeq 0.964 ,\; r\simeq
0.051 $. This value for $ r $ is within reach of forthcoming CMB
observations. }

\medskip

\item {{\bf(d)} The present data in the effective theory of
inflation clearly {\bf prefer new inflation}. }

\medskip

\item { {\bf(e)} Study of higher degree
inflaton potentials show that terms of degree higher than four do not
affect the fit in a significant way. In addition, horizon exit happens for
$ \varphi/[\sqrt{N} \; M_{Pl}] \sim 0.9 $ making higher order terms
in the potential $ w $ negligible. }

\end {itemize}

\bigskip

We summarize the physical effects of
{\bf generic} initial conditions (different from Bunch-Davies) on the
scalar and tensor perturbations during slow-roll and
introduce the transfer function $ D(k) $ which encodes the observable
initial conditions effects on the power spectra.
These effects are more prominent in the \emph{low}
CMB multipoles: a change in the initial conditions during slow roll can
account for the observed CMB {\bf quadrupole suppression}.

\bigskip

Slow-roll inflation is generically preceded by a
short {\bf fast-roll} stage. Bunch-Davies initial conditions are the
natural initial conditions for the fast-roll perturbations. During
fast-roll, the potential in the wave equations of curvature and tensor
perturbations is purely attractive and leads to a suppression of the
curvature and tensor CMB quadrupoles. 

\bigskip

A MCMC analysis of the WMAP+SDSS data {\bf including fast-roll} shows that the quadrupole
mode exits the horizon about 0.2 efold before fast-roll ends and its
amplitude gets suppressed. In addition, fast-roll fixes the {\bf initial
inflation redshift} to be $ z_{init} = 0.9 \times 10^{56} $ and
the {\bf total number} of efolds of inflation to be $ N_{tot} \simeq 64 $.

\bigskip

Fast-roll fits the TT, the TE and the EE
modes well reproducing the quadrupole supression. 

\bigskip

A thorough study of the
{\bf quantum loop corrections} reveals that they are very small and controlled by
powers of $(H /M_{Pl})^2 \sim {10}^{-9} $, {\bf a conclusion that validates the
reliability of the
effective theory of inflation.} 

\bigskip

The present review shows how powerful is
the Ginsburg-Landau effective theory of inflation in predicting
observables that are being or will soon be contrasted to observations.

\bigskip

\bigskip

\begin{center}

{\bf References, Review article: 

\bigskip

\noindent
D. Boyanovsky, C. Destri, H. J. de Vega, N. G. Sanchez,

\medskip

\noindent

arXiv:0901.0549, Int. J. Mod. Phys. A24, 3669-3864 (2009)

\medskip

and  author's references therein.}

\end{center}

\newpage

\subsection{H.J. de Vega,  N.G. Sanchez }

\vspace{-0.3cm}

\begin{center}

H.J.dV: LPTHE, CNRS/Universit\'e Paris VI-P. \& M. Curie \& Observatoire de Paris, Paris, France\\
N.G.S: LERMA, CNRS/Observatoire de Paris, Paris, France

\bigskip

{\bf Dark Matter at the  keV scale from Theory and Observations}

\end{center}

\bigskip

The nature of Dark Matter (DM) is unknown. It is a forefront problem of modern cosmology. Only the gravitational effects of DM are observed and they are necessary to explain the present structure of the Universe in the context of the standard Cosmological model. DM particles must be neutral and so weakly interacting that no effects are so far detectable. There are extremely many DM particle candidates beyond the standard Model of particle physics.

\bigskip

A new analysis of the dark matter particle mass, taking into account theory, galaxy observations and numerical simulations indicates that the mass of the dark matter particle is in the keV scale  (keV = $1/511$ electron mass) and the temperature when the dark matter  decoupled from ordinary matter and radiation would be $100 GeV$ at least. [1-4]. 

\bigskip

This analysis is based on  the generic properties of the distribution function and the phase density of dark matter particles,  {\bf independent of the particle physics model}. The several generic possibilities for the dark matter particles have been considered: at  decoupling they could be ultra-relativistic or non-relativistic, at or out local thermal equilibrium. In all cases, the dark matter  particles are "cold" enough to allow galaxy formation, their mass turns to be at the keV scale and the dark matter interactions (other than gravity)  are negligible.

\bigskip

[So far, the search for dark matter particles concentrated unsuccessfully on much heavier particles with masses of  10 GeV or more ].

\bigskip

Two independent constraints are used:  The known cosmological DM density today $\rho_{DM}({\rm today})= 1.107 \; \frac{\rm keV}{{\rm cm}^3}$
,   and the phase-space density $Q=\rho/\sigma^3$ which is invariant under the cosmological expansion and can only decrease under self-gravity interactions (gravitational clustering). The value of $ Q $ today follows from galaxy observations :
$ Q_{today} = (0.18 \;  \mathrm{keV})^4 $. 
We compute explicitly $ Q_{prim} $ (in the primordial universe) and it turns
to be proportional to $ m^4 $ [1-4].

\bigskip

Alternatively, we use the surface acceleration of gravity in DM dominated galaxies and thus provide two quantitative ways to derive the value of  $m $ and the decoupling temperature $T_d$ in refs. [1-4]. The dark matter particle mass  $ m $ and decoupling temperature 
$ T_d $ are {\bf mildly} affected by the uncertainty in the factor $ Z $ through a power factor 
$ 1/4 $ of this uncertainty, namely, by a factor $ 10^{\frac14} \simeq 1.8 $

\bigskip

No assumption about the nature of the dark matter 
particle is made. The keV range DM particle mass is much larger than the temperature 
during the matter dominated era (which is less than 1 eV), hence the keV dark matter 
is {\bf cold} (CDM). 

\bigskip

The comoving Jeans' (free-streaming) wavelength, ie the largest wavevector 
exhibiting gravitational instability  (Fig. 1), and the Jeans' mass (the smallest 
unstable mass by gravitational collapse) are obtained in the range
$$
\frac{0.76}{\sqrt{1+z}} \; {\rm kpc} <\lambda_{fs}(z) <
\frac{16.3}{\sqrt{1+z}} \; {\rm kpc} \; , \; 0.45 \; 10^3 \; M_{\odot} 
< \frac{M_J(z)}{(1+z)^{+\frac32}} < 0.45 \; 10^7  \; 
\; M_{\odot} \; .
$$

These values at $ z = 0 $ are consistent with the $N$-body simulations 
and are of the order of the small dark matter structures observed today .
By the beginning of the matter dominated era $ z \sim 3200 $, the masses are of the 
order of galactic masses $ \sim 10^{12} \; M_{\odot} $ and the comoving free-streaming 
wavelength scale turns to be of the order of the galaxy sizes today $ \sim 100 \; {\rm kpc}  \; ,$.

\bigskip

 Lower and upper bounds for the dark matter annihilation cross-section $ \sigma_0 $ 
are derived: $\sigma_0 > (0.239-0.956) \; 10^{-9} \; \mathrm{GeV}^{-2}$ and 
$\sigma_0 < 3200 \; m \; \mathrm{GeV}^{-3} \; .$ There is at least five orders of 
magnitude between them , the dark matter non-gravitational self-interaction is 
therefore negligible (consistent with structure formation and observations, as 
well as by comparing X-ray, optical and lensing observations of the merging of 
galaxy clusters with $N$-body simulations).

\bigskip

Typical "wimps" (weakly interacting massive particles) with mass $ m = 100 $ GeV 
and $ T_d = 5 $ GeV  would require a huge $ Z \sim 10^{23} $, well above
the upper bounds obtained and cannot reproduce the observed galaxy properties. 
They produce an extremely short free-streaming or Jeans length $ \lambda_{fs} $ today $ 
\lambda_{fs}(0) \sim 3.51 \; 10^{-4} \; {\rm pc} = 72.4  \; {\rm AU} \; $ that would
correspond to unobserved structures much smaller than the galaxy structure.
Wimps result strongly disfavoured. 

\bigskip

{\bf References}

\begin{description}

\bigskip

\item[{\bf [1]}]  H. J. de Vega, N. G. Sanchez,  arXiv:0901.0922,
Mon. Not. R. Astron. Soc. 404, 885 (2010).

\medskip

\item[{\bf [2]}] D. Boyanovsky, H. J. de Vega, N. G. Sanchez,
arXiv:0710.5180, Phys. Rev. D77, 043518 (2008)

\medskip

\item[{\bf [3]}] H. J. de Vega, N. G. Sanchez, arXiv:0907.0006.

\medskip

\item[{\bf [4]}] D. Boyanovsky, H. J. de Vega, N. G. Sanchez,
arXiv:0807.0622, Phys. Rev. D78, 063546 (2008).

\end{description}

\newpage

\subsection{Massimo Giovannini}

\centerline{ Department of Physics, Theory Division, CERN, 1211 Geneva 23, Switzerland} 

\medskip

\centerline{INFN, Section of Milan-Bicocca, 20126 Milan, Italy}

\bigskip

\begin{center}

{\bf Large-scale magnetic fields in the standard  model }

\end{center}
 
\bigskip

For reasons of space, only some of the  results obtained during 2009 have been reported in this talk. Relevant references are:

\medskip

\begin {itemize}

\item {(1) M. Giovannini,   Phys. Rev. D79, 121302 (2009).}
\item {(2) M. Giovannini,  Phys. Rev. D79, 103007 (2009).}
\item {(3) M. Giovannini and N. Q. Lan, Phys. Rev. D80 027302(2009).}
\item {(4) M. Giovannini and K. Kunze   Phys. Rev. D79, 063007 (2009).}
\item {(5) M. Giovannini, CERN-PH-TH-2009-117, arXiv:0907.3235 [astro-ph.CO].}

\end {itemize}

\medskip

 The parameters of a putative magnetized background have been estimated, for the first time, from the observed temperature autocorrelation (TT angular power spectra) as well as from the measured temperature-polarization cross correlation (TE angular power spectra) (see (1)-(2)).  

\medskip

{\bf Likelihood contours have been presented} (see (1)-(2)).  The dependence of the temperature and polarization angular power spectra upon the parameters of an ambient magnetic field can be encoded in the scaling properties of a set of basic integrals whose derivation is simplified in the limit of small angular scales. 

\medskip

The magnetically-induced distortions patterns of the relevant observables can be computed analytically by employing scaling considerations which are corroborated by numerical results. The parameter space of the magnetized cosmic microwave background anisotropies is also discussed in the light of the obtained analytical results (see (2)-(3)). The propagation of electromagnetic disturbances in a magnetized plasma leads naturally to a B-mode polarization whose angular power spectrum can be computed both analytically and numerically (4).  

\medskip

A strategy for the direct extraction of the magnetized B-mode autocorrelations from the forthcoming experimental data has been presented and discussed.  Taken at face value, the results presented here and reported in the aforementioned publications,  illustrate, for the first time, how the parameters of a magnetized background can be systematically  included and estimated in the LambdaCDM paradigm as well as in its neighboring extensions (5).

\medskip

{\bf The research program illustrated in this talk has been formulated through various steps and the references quoted in (1)-(5) can be usefully consulted}.

\newpage

\subsection{A. Sasha Kashlinsky}

\medskip

\centerline {NASA GSFC: Goddard Space Flight Center, Greenbelt, MD, USA }

\centerline {with F. Atrio-Barandela (Salamanca, Spain),
D. Kocevski (UCDavis),
H. Ebeling (U Hawaii)}

\medskip

\bigskip

\centerline {\bf Large-scale peculiar flows of clusters of galaxies}

\bigskip

\medskip

In the standard cosmological paradigm, large-scale peculiar velocities arise from
gravitational instability due to mass inhomogeneities seeded during inflationary
expansion. On sufficiently large scales, > 100 Mpc, this leads to a robust prediction of the amplitude and coherence length of these velocities. 

\bigskip

For clusters of galaxies, their peculiar velocities can be measured from the kinematic component of the Sunyaev-Zeldovich (SZ) effect produced by the Compton scattering of cosmic microwave background (CMB) photons off the hot intracluster gas. This talk discusses results from measurements of the large scale peculiar flows using a large X-ray cluster catalog and all sky CMB maps from the WMAP satellite (Kashlinsky et al 2008, ApJ, 686, L49 and 2009, 691, 1479). 

\bigskip

The analysis utilizes the method proposed by us earlier (Kashlinsky and Atrio-Barandela 2000, ApJ, 536, L67): it computes the dipole in the cosmic microwave (CMB) data at cluster pixels, which preserves the KSZ component, while integrating down other contributions. 

\bigskip

In a parallel study we demonstrated that the hot gas in clusters is well described by the Navarro-Frenk-White density profile (Atrio-Barandela et al 2008, 675, L57). Such NFW clusters have gas temperature decrease toward outer parts consistent with the available X-ray measurements, so the thermal SZ integrates down with increasing cluster aperture enabling to isolate the KSZ component in the dipole. 

\bigskip

The discussion addresses in great detail the possible systematics that can confuse our measurements and it is demonstrated that - given the quality of the cluster catalog - the various systematic effects are small and cannot reproduce the measured dipole.

\bigskip

\subsection{ Eiichiro Komatsu}

\vskip -0.5cm

\begin{center}

University of Texas, Department of Astronomy, Austin, USA.

\bigskip

\medskip

{\bf How WMAP Helps Constrain the Nature of Dark Energy} 

\end{center}

\bigskip

I presented a method to compress the information on the nature of dark  
energy, contained in the cosmic microwave background (CMB) data  
obtained by the WMAP satellite, to just three numbers, and showed the  
latest limits on time evolution of dark energy density using the  
distance information (angular diameter distances from CMB and the  
distribution of galaxies, as well as luminosity distances from Type Ia  
supernovae) alone. 

\bigskip

The current limits are fully consistent with dark  
energy being a cosmological constant, with the present-day equation of  
state parameter constrained as $w=-1.00\pm 0.19$ (68\%CL). 

\bigskip

I then presented a way to improve on this limit significantly, by including  
the full power spectrum information contained in the distribution of  
galaxies. 

\bigskip

As an example I presented how the Hobby-Eberly Dark Energy  
Experiment (HETDEX), which is a large redshift survey developed by the  
University of Texas and the partner institutions, can determine the  
important quantities for constraining the nature of dark energy - the  
angular diameter distances and Hubble expansion rates - with much  
better (more than a factor of two) accuracy compared with a now- popular method that uses only the Baryon Acoustic Oscillations. The  
HETDEX survey is scheduled to begin in 2011.

\bigskip

{\bf References:}

\begin{description} 
\item{1} Komatsu et al., ApJS, 180, 330-376 (2009)
\vskip -0.2cm
\item{2}Shoji, Jeong and Komatsu, ApJ, 693, 1404-1416 (2009)

\end{description}

\newpage

\subsection{ Anthony Lasenby}

\vskip -0.2cm

\begin{center}

 Astrophysics Group, Cavendish Laboratory, J.J. Thomson Avenue,
Cambridge CB3 0HE, U.K. and Kavli Institute for Cosmology, c/o Institute of 
Astronomy, Madingley Road, Cambridge, CB3 0HA, U.K. \\
                                       Email: a.n.lasenby@mrao.cam.ac.uk

\end{center}

\bigskip

\centerline{{\bf CMB Observations: Current Status and Implications for Theory}}

\bigskip
           
\medskip
                                
The Cosmic Microwave Background (CMB), is a wonderful tool in modern
cosmology. A significant fraction of all the information in cosmology over the last 10 to 15 years has come from it, and it has finally ushered us into an era of 'precision cosmology' (the latter, of course, accompanied by deep mysteries
as to the nature of the quantities, such as 'dark energy' and 'dark matter', which we are measuring so accurately). 

\bigskip

The aim of the talk was to give an overview of the current state of CMB observations and their scientific implications.

\bigskip

One of the big questions that current and forthcoming CMB observations can
help with, is the dynamics and energy scale of inflation. A key observation in this respect would be detection of B-mode CMB polarization, which would enable us to determine the parameter r, the ratio of tensor to scalar modes of primordial perturbations. In this respect, some new interesting polarization results are coming out from two current experiments.

\bigskip

Recent results from the QUAD experiment at the South Pole ([1]) show that
the expected peak structure in the E-mode at scales between about 200 to 2000 in l  has been definitely detected, at high significance.

\bigskip

Recent results from BICEP (also at the South Pole), give a direct limit to
the B mode level of $ r < 0.73$ at 95\% confidence ([2]). This is much larger
than the limit of $r < 0.22$ at 95\% given by [4]. However, the latter is not a direct limit, but comes via a combination of constraints from T and E mode CMB (principally from the 5 year WMAP observations), together with large scale
structure data and supernovae. Chiang et al. [2] show that the direct upper limit on B-modes from current WMAP data is $r < 6$.

\bigskip

The next two years should see a very considerable improvement in CMB
measurements on all scales with data starting to come through from the Planck satellite, which was successfully launched on May 14th 2009. Detection of an $r$ value as low as $0.05$ should be possible with a Planck mission that includes 4
sky coverages (see [3]), as well as a much improved measurement of the slope of the primordial scalar spectrum, $n_s$ . 

\bigskip

Also vitally important for discriminating between competing theories of inflation, are $n-run$ (i.e. is the slope of the primordial spectrum fixed, or does it change with wavenumber), and the question of whether the primordial fluctuations are Gaussian. It is now clear that
estimators like $f_nl$ are very good discriminators of the type of inflation, and Planck should give at least a fourfold improvement in accuracy of measurement for this quantity.

\bigskip

Similar advances are also being made for secondary anisotropies. The first
'blank field' Sunyaev-Zeldovich detections have appeared from the South Pole Telescope [6], and the talk included images from the AMI telescope in Cambridge ([7]) which is now well into its first deep blank-field survey. 

\bigskip

Other areas in which the CMB can provide crucial information include topics in
'fundamental physics', such as possible constraints (via string cosmology) on the nature of quantum gravity, the detection of topological defects,
and the question of whether the universe may have been non-isotropic when
the perturbations on largest scales were being laid down. For more on these and related topics see [5].

\bigskip

{\bf References}

\bigskip

\begin{description}

\item[{\bf[1]}] M. L. Brown, et al. Improved measurements of the temperature and
polarization of the CMB from
QUaD. ArXiv e-prints, June 2009. arXiv:0906.1003.

\medskip

\item[{\bf[2]}]H. C. Chiang, et al. Measurement of CMB Polarization Power Spectra
from Two Years of BICEP
Data. ArXiv e-prints, June 2009. arXiv:0906.1181.

\medskip

\item [{\bf[3]}] G. Efstathiou and S. Gratton. B-mode detection with an extended planck mission. Journal of
Cosmology and Astro-Particle Physics, 6:11-+, June 2009.

\medskip

\item [{\bf[4]}] E. Komatsu, et al. Five-Year Wilkinson Microwave Anisotropy Probe Observations: Cosmological
Interpretation. ApJS, 180:330-376, February 2009.

\medskip

\item [{\bf[5]}] A.N. Lasenby. The Cosmic Microwave Background and Fundamental Physics. Space Science Reviews,
online version doi:10.1007/s11214-009-9616-4, 2010.

\medskip

\item [{\bf[6]}] Z. Staniszewski, et al. Galaxy Clusters Discovered with a Sunyaev-Zel'dovich E-ect Survey. ApJ, 701:32-41, August 2009.

\medskip

\item [{\bf[7]}] J. T. L. Zwart and AMI Consortium. The Arcminute Microkelvin Imager. MNRAS, 391:1545- 1558, December 2008.

\end{description}

\newpage

\section{Summary and Conclusions of the Colloquium by H.J. de Vega, M.C. Falvella and N.G. Sanchez}

\bigskip

About one hundred participants (from Europe, North and South America, Japan, Russia, Armenia, Latvia, India, Korea, Taiwan, New Zealand, South Africa ) attended the Colloquium.

\bigskip

All  the announced 19 Lecturers were present, including 4 Daniel Chalonge Medals, among them George Smoot, Nobel prize of Physics. News from WMAP and from Planck were directly reported. Journalists and representatives of the directorate of the Italian Space Agency were present.

\bigskip

Discussions and lectures were outstanding. Inflection points in several current research lines emerged.  New important issues and conclusions arised and between them, it worths to highlight:

\bigskip

\begin{itemize}

\bigskip

\item{ {\bf(1)} The primordial CMB fluctuations are almost gaussian, large primordial non-gaussianity  and large primordial running index are strongly disfavored. The amount of primordial gravitons r  is predicted to be larger than 0.021 and smaller than 0.053, which is at reach of the next CMB observations.}

\bigskip

\item{{\bf(2)} The dark matter particle candidates with high mass (100 GeV, the so called "Wimps") became strongly disfavored, while cored (non cusped) dark matter halos  and light (keV scale mass) dark matter are being increasingly favoured from theory and  astrophysical observations.}

\bigskip

\item{{\bf(3)} Dark energy observations are pretty consistent with the cosmological constant. CMB + BAO is the winner for measuring spatial curvature, but other standard rulers are to be considered beyond BAO as the horizon size at the matter-radiation equality era, z ~3200. The HETDEX survey is expected to determine important quanties for constraining dark energy -as the angular diameter distances and Hubble expansion rates - with much better (more than a factor of two) accuracy compared with methods  using only  BAO. HETDEX is scheduled to begin in 2011.}

\bigskip

\item{{\bf(4)} The features of electrons and positrons observed recently by Auger, Pamela and HESS are all explained as having their origin  in the explosions and winds of massive stars in the Milky Way.}

\bigskip

\item{{\bf(5)} The QUAD experiment at the South Pole shows that the expected peak structure in the E-mode CMB polarization at scales between about 200 to 2000 in l  has been definitely detected  at high significance. Detection of an r value as low as 0.05 should be possible with a Planck mission that includes 4 sky coverages, as well as a much improved measurement of the scalar primordial index, ns. The first sky strips (first light survey) observed by Planck launched 14th may were presented in avant prèmière.}

\bigskip

\item{{\bf(6)}  Advances are also being made for  CMB secondary anisotropies.The first 'blank field' Sunyaev-Zeldovich detections have appeared from the South Pole Telescope, and new images were presented  from the AMI telescope in Cambridge which is now well into its first deep blank-field survey.}

\end{itemize}
  
\bigskip
  
Best congratulations and aknowledgements to all  lectures and participants which made the 13th Paris Cosmology Colloquium so fruitful and interesting, the Ecole d'Astrophysique Daniel Chalonge looks forward for you for the next Colloquium of this series.

\newpage

\section{Award of the Daniel Chalonge Medal 2009}

\bigskip

The International School of Astrophysics "Daniel Chalonge" has awarded the Daniel Chalonge Medal 2009 to {\bf Professor Peter Biermann} from the MPI for Radioastronomie of Bonn (D) and University of Alabama-Tuscaloosa (USA).

\bigskip

The medal was awarded to Peter Biermann for his pioneering, impressive and multiple contributions to astrophysics (as for example high energy particle acceleration, cosmic rays, galactic nuclei and black holes), and for his support and outstanding contributions to the Chalonge School. In particular, Peter Biermann is involved in astrophysical dark matter research in the standard model of the universe, one of the most discussed topics in the Chalonge School.  Peter Biermann takes part in  the programs and life of the School, promoting fruitful discussions and work with the participants and supporting the origin and development of new ideas and projects.

\bigskip

The Chalonge medal was presented to Peter Biermann on July 25, 2009 during the sessions of the 13th Paris Cosmology Colloquium 2009 at the Observatoire de Paris HQ (historic Perrault building) in the Cassini Hall, on the meridian of Paris, which was attended by about hundred participants from the world over, among them three Chalonge Medals.

\bigskip 

The Chalonge Medal, coined exclusively for the Chalonge School by the prestigious Hotel de la Monnaie de Paris (the French Mint), is a totally surprise award and only seven Chalonge medals have been awarded in the 18 year school history. 

\bigskip

The Medal aknowledges science with great intellectual endeavour and a human face. True and healthy science. Outstanding gentleperson scientists. Scientists recipients of the Daniel Chalonge Medal are Ambassadors of the School.

\begin{figure}[htbp]
\epsfig{file=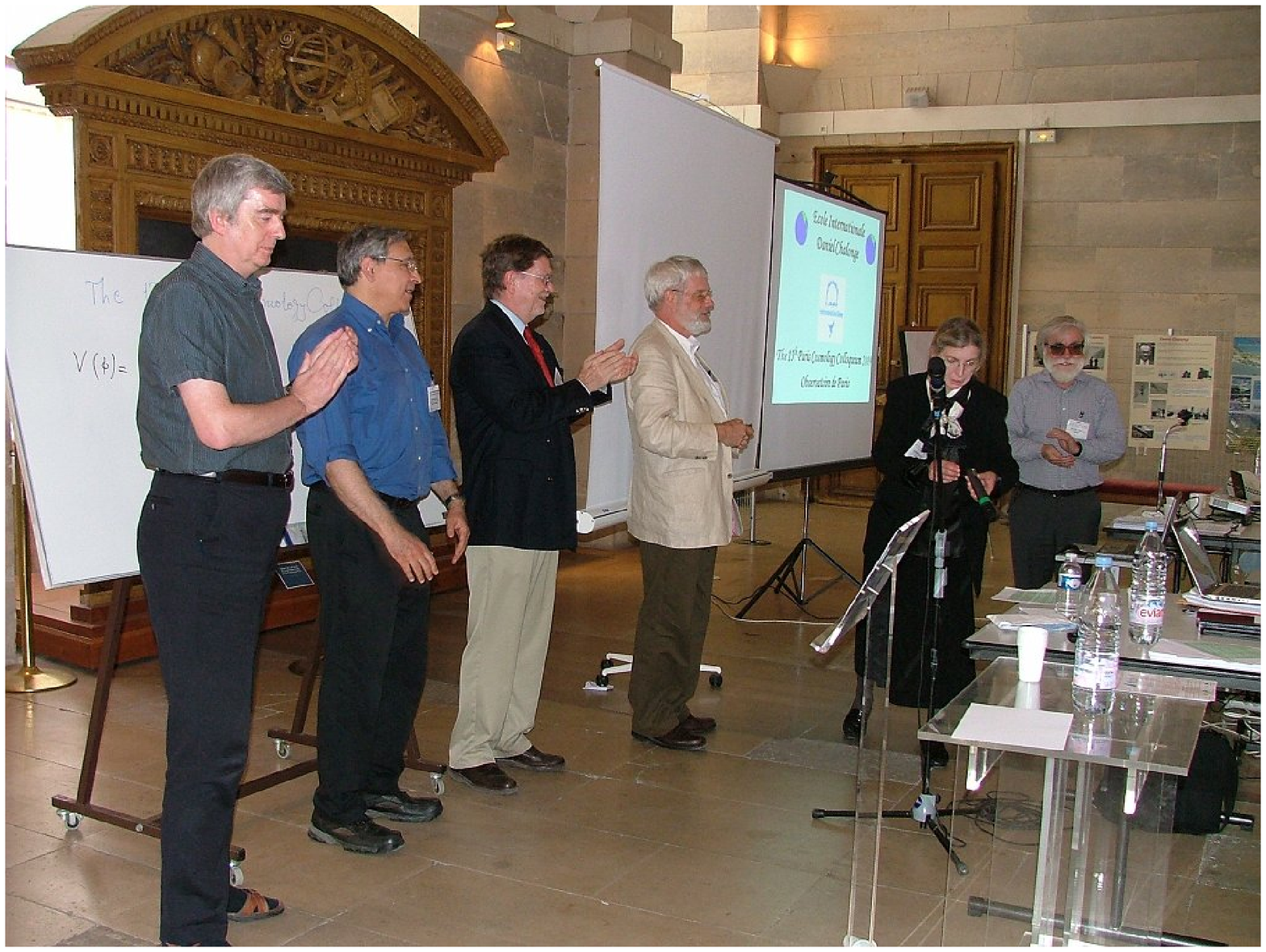,width=16cm,height=12cm}
\end{figure}

\bigskip

The list of the awarded Chalonge Medals is the following:

\bigskip

\begin {itemize}

\item {1991: Subramanyan Chandrasekhar, Nobel prize of physics.}
\item {1992: Bruno Pontecorvo.}
\item {2006: George Smoot, Nobel prize of physics.}
\item {2007:  Carlos Frenk}
\item {2008: Anthony Lasenby}
\item {2008: Bernard Sadoulet.}
\item {2009: Peter Biermann.}

\end{itemize}

\bigskip

The announcement, full history, photo gallery and links are available on line at:

\bigskip

\begin {center} 

{\bf http://chalonge.obspm.fr}  ~ , ~ click on `The Daniel Chalonge Medal 2009'

\bigskip

{\bf http://chalonge.obspm.fr/Medal\_Chalonge2009.pdf}

\end {center}

\bigskip

\begin {center}

PHOTOS OF THE COLLOQUIUM

\bigskip

are available at `Album photos': 

\bigskip

{\bf http://chalonge.obspm.fr/colloque2009.html}

\end {center}

\newpage

\section{List of Participants}

\bigskip

ABEL 	Thomas, Stanford University, Physics Department, Stanford, California,	USA\\
\medskip
Mrs ABEL,  Stanford, California,	USA\\
\medskip
ABREU  Gabriel, Victoria University of Wellington, Wellington,  NEW  ZEALAND\\
\medskip
AFANASIEV   Mikhail,	Space  Research Institute,	Moscow,	RUSSIA\\
\medskip
AMES	  Susan,	 Oxford  University ,  Dept of Astrophysics,  Oxford,  ENGLAND\\
\medskip
ASADA  Hideki,	Hirosaki University,  Hirosaki,	JAPAN\\	
\medskip
BAACKE   Jürge,  Fachbereich Physik,  Dortmund University, Dortmund, GERMANY\\
\medskip
BARTOLO   Nicola,	Università di  Padova, Dipt di Fisica "Galileo Galilei",  Padova, ITALY\\	
\medskip
BASAK	Soumen,	Institut d'Astrophysique de Paris,   Paris,	FRANCE\\
\medskip
BAUMONT	Sylvain,	LPSC-IN2P3-CNRS,	  Grenoble,	FRANCE\\
\medskip
BECKER Ulrich,	Massachusetts Institute of Technology, Cambridge/ MA, USA\\
\medskip
BIERMANN	Peter L. , MPI-Bonn and Univ of Alabama-Tuscaloosa, GERMANY, \\
\medskip 
BONOMETTO  Silvio,    Univ Milano-Bicocca, Dipt di Fisica,	Milano, ITALY	\\
\medskip
BULLOCK	James,	 University of California at Irvine, Physics and Astr, Irvine, CA, USA	\\
\medskip
CAO	Francisco J.,Universidad Complutense de Madrid, Dept Fisica Atom., Madrid,  SPAIN\\
\medskip
CHATTERJEE  Sujit, Relativity and Cosmology Center,  Jadavpur Univ, Kolkata, INDIA	\\
\medskip
CLERC	Nicolas,	CEA/Saclay,  IRFU/Sap,  Saclay,	FRANCE\\
\medskip
CLINE  David, UCLA, University of California at Los Angeles, , Los Angeles CA,USA	\\
\medskip
CNUDDE  Sylvain,  LESIA Observatoire de Paris,  Meudon,  France \\
\medskip
COORAY	Asantha, University of California at Irvine, Phys and Astr., Irvine, CA,	USA\\
\medskip
DAGTEKIN  Nazli D, Erciyes University, Radio Astronomy Observatory, Kayseri, TURKEY\\
\medskip
DAVAL	Benoit	,   Paris	France	,   FRANCE\\
\medskip
DEBNATH	Ujjal,	Bengal Engineering and Science University, Howrah,  INDIA	\\
\medskip
DEBONO	Ivan,	Service d'Astrophysique,  CEA Saclay,  Paris,  FRANCE\\
\medskip
DECHANT	Pierre-Philippe,  Cambridge University, Cambridge,  UNITED KINGDOM	\\
\medskip
DEMOCLES	 Jessica,	IRFU-CEA-Saclay,	Saclay,  FRANCE\\
\medskip
DESTRI  Claudio, Univ Milano-Bicocca /INFN, Dipt di Fisica G. Occhialini, Milano, ITALY	\\
\medskip
DE VEGA   Héctor,  Universté Pierre et Marie Curie LPTHE and CNRS,  Paris,  FRANCE\\
\medskip
DING	Ran, 	 Shanghai  Normal  University,    Shanghai, 	  CHINA\\
\medskip
DOMINGUEZ   Mariano,  IATE-OAC,  Observatorio Astronomico , Cordoba, ARGENTINA\\
\medskip
DVOEGLAZOV     Valeriy,  Universidad de Zacatecas,  Zacatecas,  MEXICO	\\
\medskip
ECHAURREN  Juan,	Codelco Chile - North Division, Electronic Laboratory, Calama, CHILE	\\
\medskip
ERDOGDU  Pirin, University College London/American University , London, UK\\
\medskip
FALVELLA	Maria Cristina,  Italian Space Agency and Univ of Roma I , Rome, 	ITALY	\\
\medskip
FELDMAN	Hume A. , University of Kansas, Cosmology Group,  Lawrence, Kansas, 	USA\\
\medskip
FONTE   Roberto,  Inst Nazionale di Fisica Nucleare-Sezione di Catania, Catania,  ITALY\\
\medskip
FOUKZON	Jaykov,     Israel  Institute  of  Technology,   Tel-Aviv,   ISRAEL\\
\medskip
FRENK  Carlos S.,  Computational Cosmology Center,  Univ.of  Durham, Durham, UK	\\
\medskip
Mrs  Susan  FRENK,   Durham,   UK\\
\medskip
GALBANY	Lluís,	Institut de Física d'Altes Energies (IFAE), Barcelona,  SPAIN\\
\medskip
GAN	Jianling, Max Planck Institute for Astronomy, Heidelberg,  GERMANY	\\
\medskip
GERGELY	Laszlo,	University of Szeged	, Szeged,	HUNGARY\\
\medskip
GHOSH  Subir,  Indian  Statistical Institute,  Kolkata, 	INDIA\\
\medskip
GILMORE	Gerard, Institute of Astronomy, Madingley Road,	Cambridge,	UK	\\
\medskip
GIOCOLI	Carlo,	IZAH, ITA - University of Heidelberg, Heidelberg, GERMANY	\\
\medskip
GIOVANNINI  Massimo, INFN-Univ. Milano-Bicoccca, Dipt di Fisica, Milano,	 ITALY \\	
\medskip
GOHEER 	Naureen,  University of Cape Town ,  Cape Town,   SOUTH AFRICA \\
\medskip
GOLBIAK  Jacek,  Catholic University of Lublin, Dept of Theor.Physics, Lublin, POLAND	\\
\medskip
GOMES  Jean Michel,  Observatoire de Paris, GEPI,  Meudon,   FRANCE	\\
\medskip
GU   Je-An Leung,  Center for Cosmology and Particle Astrophysics, Taipei, TAIWAN \\
\medskip
GUPTA Rajiv,  Physics Department, Guru Nanak Dev University, Amritsar, INDIA	\\
\medskip
HANZEVACK Emil,	   College of  William  and  Mary,  Williamsburg, Virginia, USA	\\
\medskip
HARUTYUNYAN  Gohar,	Yerevan State University  YSU, Yerevan,	ARMENIA	\\
\medskip
HILDEBRANDT  Sergi,  Instituto de Astrofisica de Canarias, La Laguna, Tenerife, SPAIN	\\
\medskip
HOST  Ole,  Dark Cosmology Centre,   Niels Bohr Institute,   Copenhagen,   DENMARK \\
\medskip
ILIEV	 Ilian,	University of Zurich,  Zurich,  SWITZERLAND 	\\
\medskip
JASNIEWICZ	 Gérard, GRAAL Université Montpellier 2 / CNRS, Montpellier,	FRANCE\\
\medskip
KAO   W.F.,	 Institute of Physics,     Chiao Tung University, 	Hsin Chu,	TAIWAN \\
\medskip
KARCZEWSKA  Danuta,   University of  Silesia,  Katowice, 	POLAND \\
\medskip
KASHLINSKY  Alexander, 	NASA  Goddard  Space Flight Center, Greenbelt,  MD,USA	\\
\medskip
KHADEKAR	Goverdhan,	RTM  Nagpur  University,   Nagpur,	  INDIA\\
\medskip
KOMATSU	Eiichiro, University of Texas at Austin, Dept of Astronomy, Austin, TX,	USA\\
\medskip
Mrs  KOMATSU, Austin,  Texas,  USA\\
\medskip
KONTUSH	Anatol, Université Pierre et Marie Curie UPMC  Paris 6, Paris, FRANCE \\
\medskip
KOSTRO	Ludwik, University of Gdañsk,  Gdansk,  POLAND \\
\medskip
KRAWIEC	Adam, 	 Jagiellonian  University,  Krakow,	POLAND\\
\medskip
KUMAR 	Jaswant,  Harish Chandra Research Institute, Allahabad,  INDIA\\	
\medskip
LALOUM	Maurice,   CNRS / IN2P3, Paris LPNHE- Jussieu,  Paris,  FRANCE\\	
\medskip
LARSEN  Arne Lykke,  University of Southern Denmark, Physics Dept.Odense, DENMARK \\ 
PEDERSEN    Stephan Klimt,     Assistant,     Odense,     DENMARK  \\
NIELSEN       Mai Drost ,          Assistant,       Odense,     DENMARK \\ 
SANKO         Cecilie,               Assistant,         Odense,     DENMARK  \\
\medskip
LASENBY	Anthony, Cavendish Laboratory, Astrophysics, Univ of Cambridge, UK\\	
\medskip
Mrs LASENBY, Cambridge, UNITED KINGDOM  \\
\medskip
LE  GOFF	Jean-Marc,	CEA Saclay,  Saclay, FRANCE\\
\medskip
LEE	Wolung,   National Taiwan Normal University,  Taipei,	TAIWAN \\
\medskip
LETOURNEUR   Nicole,	Observatoire de Paris LESIA  Meudon,  FRANCE \\	
\medskip
LOPES	Paulo,	IPD / Univap,	São José ,	BRAZIL \\
\medskip
MACHADO	André,  Fundacao FCUL,  Lisbon  University, Lisbon,  PORTUGAL	\\
\medskip
MAIO	  Umberto,	Max  Planck  Institute,    Garching,   GERMANY \\
\medskip
MANDOLESI   Reno, 	IASF-Bologna, INAF	,	Bologna,	ITALY	\\
\medskip
Mrs   MANDOLESI,    Bologna,     ITALY \\
\medskip
MATARRESE  Sabino, Università di Padova, Dipt di Fisica "Galileo Galilei",Padova,ITALY \\
\medskip
MARTI    Pol,	     Institut de Física d'Altes Energies (IFAE),   Barcelona,   SPAIN \\
\medskip	
MATELOT-LECROSNIER Nathalie, Senior Comm. Coord., EDP Sciences, Paris,  FRANCE \\
\medskip
MATHEWS Grant,  University of Notre Dame, Center for Astrophysics,  Notre Dame,	USA	\\
\medskip
MAZUMDAR Anupam, Lancaster University and Copenhagen University,Lancaster,	UK	\\
\medskip
MAZURE	Alain,	 LAM/CNRS, OAMP  Marseille,  FRANCE	\\
\medskip
MEDARI    Leila,   Faculté des Sciences Physiques - Université Caddi,  Marrakech,  MAROC \\
\medskip
MERSINI-HOUGHTON  Laura, Univ.of North Carolina-Chapel Hill, USA and DAMTP,UK.\\
\medskip
MICKAELIAN Areg,	 Byurakan Astrophysical Observatory BAO, Yerevan,  ARMENIA	\\
\medskip
MISKIN Vitthal, Yahvantrao Chuhan College of Engineering (YCCE), Nagpur,	INDIA\\
\medskip
MYCHELKIN  Eduard G. ,  Nat. Center of Space, Astrophys. Inst,  Almaty, KASAKHSTAN\\
\medskip
NAGAI	Daisuke,	Yale University,  Physics Dept.,  New Haven,	USA	\\
\medskip
NOH	Hyerim, Korea Astronomy and Space Sceince Institute, Taejon, 	KOREA\\
\medskip	
OVGUN	Ali,	Izmir Institute of Technology,  Izmir, 	TURKEY\\
\medskip
PANDYA 	Aalok,	Department of  Physics, University of Rajasthan, Jaipur,	INDIA\\
\medskip
PEÑA SUAREZ Vladimir Jearim,Univ. Industrial de Santander, Bucaramanga,	COLOMBIA\\
\medskip
PFEIFER   Anna,   Bonn,   GERMANY \\
\medskip
PFEIFER    Monika , Bonn,    GERMANY\\
\medskip	
PILOYAN	Arpine,	Yerevan State University, Yerevan,	ARMENIA	\\
\medskip
RACCANELLI Alvise, Institute of Cosmology and Gravitation,	Portsmouth,	UK\\
\medskip
RAETH  Christoph,	Max-Planck Institute, Garching, GERMANY \\	
\medskip
RAMON  MEDRANO Marina, Univ.Complutense de Madrid, Fisica Teor I, Madrid, SPAIN\\
\medskip
RATCLIFFE   Kathy,   De Montford University,  Leicester,  ENGLAND\\
\medskip
RINDLER-DALLER	  Tanja, Inst. Theoretical Physics, Univ of Cologne, Koeln, GERMANY	\\
\medskip
ROSSI  Graziano, Korea Institute for Advanced Study, Astrophysis Group,  Seoul,  KOREA\\
\medskip
SAAL   Margus,	Tartu   Observatory,	Tõravere,	ESTONIA	\\
\medskip
SALITIS	Antonijs,	Daugavpils University,  Daugavpils,	  LATVIA\\
\medskip
SALUCCI   Paolo,  SISSA,  Astrophysics  Research  Sector,  Trieste,  ITALY\\
\medskip
SANCHEZ	Norma G.	Observatoire de Paris LERMA and CNRS, 	Paris,	FRANCE	\\
\medskip
SERRA	Ana Laura,	Universita degli Studi di Torino,  Torino,	ITALY	\\
\medskip
SÉVELLEC	Aurélie,	Observatoire de Paris LESIA,  Meudon, 	FRANCE	\\
\medskip
SHAPIRO Paul R., University of Texas at Austin, Dept of Astronomy, Austin TX, USA	\\
\medskip
SKALALA	Jozef,	 Victoria University,  Wellington,	NEW ZEALAND\\
\medskip
SMOOT George F.	Lawrence Berkeley Lab.and Univ of California, Berkeley, CA, USA	\\
\medskip
SZYDLOWSKI	Marek, 	Jagiellonian University,  Krakow, 	POLAND\\
\medskip
TASINATO	Gianmassimo,	  Heidelberg  University,  Heidelberg,  GERMANY \\
\medskip
TEDESCO   Luigi,	Dipartimento di Fisica di Bari and INFN di Bari,  Bari,  ITALY\\
\medskip
TIFFENBERG   Javier,  Fac.Ciencias Exactas y Naturales, Univ Buenos Aires, ARGENTINA\\
\medskip
TONOIU  Daniel,	Institute of Space Science,	Bucharest,	ROMANIA\\
\medskip
URTADO	Olivier,  Université de Paris Sud - Orsay, Astrophysique, Orsay,  FRANCE	\\
\medskip
VALENTINI 	Antony,  Theory Group,  Imperial  College London,  London,  UK\\
\medskip
VAN DER BIJ  Jochum, Institut fuer Physik, Universitaet Freiburg, Freiburg, GERMANY	\\
\medskip
VAN ELEWYCK   Véronique,  APC-Tolbiac,  Université Paris VI,   Paris,  FRANCE\\
\medskip
VELASQUEZ  TORIB Alan Miguel, Univ. Federal de Juiz e Fora, Juiz de Fora, BRAZIL \\
\medskip
VERMA   Murli  Manohar,    Lucknow  University,   Lucknow , INDIA\\
\medskip
WANG 	Lingyu, 	University  of  Sussex, Brighton,  UNITED KINGDOM\\
\medskip
WYSE 	Rosemary,	Johns Hopkins University, Baltimore,	USA	\\
\medskip
ZANINI   Alba,   INFN Sez. di Torino and Dipt di Fisica Univ di Torino, Torino,  ITALY	\\
\medskip
ZHOGIN 	Ivan, 	Institute of Solid State Chemistry, Novosibirsk,	RUSSIA\\
\medskip
ZIDANI	Djilali,	Observatoire de Paris LERMA and CNRS,  Paris, FRANCE

\end{document}